# Current-induced effective magnetic field in a half-metallic oxide $La_{0.67}Sr_{0.33}MnO_3$

Michihiko Yamanouchi[1], Tatsuro Oyamada[2], Takayoshi Katase[1], and Hiromichi Ohta[1]

[1]Research Institute for Electronic Science, Hokkaido University, N20W10, Kita-ku, Sapporo 001-0020, Japan.

[2] Graduate School of Information Science and Technology, Hokkaido University, N14W9, Kita-ku, Sapporo 060-0814, Japan.

**Abstract**

We investigated current-induced effective magnetic field $H_{eff}$ in half-metallic oxide $La_{0.67}Sr_{0.33}MnO_3$ (LSMO) films with various thicknesses by using the planar Hall effect. Applying in-plane current to the LSMO films exerted an in-plane $H_{eff}$ orthogonal to the current direction on magnetization. The $H_{eff}$ magnitude increased with increasing current magnitude, and the direction reversed when the applied current switched to opposite sign. Assuming that a 6.5-u.c. insulating layer is created in the LSMO, the values of $H_{eff}$ observed in devices with three different LSMO thicknesses were almost scaled with current density, evaluated from the effective LSMO thickness excluding the insulating layer, suggesting that $H_{eff}$ is induced in the LSMO bulk.

Spin-orbit torques (SOTs) have attracted much attention mainly because they offer the possibility for efficient electrical manipulation of the magnetization direction in spintronics devices. SOTs originate from spin-orbit interaction at heterointerfaces (including surface) and/or in the bulk of magnetic stacking structures.[1-5] Applying in-plane current to such structures exerts an effective magnetic field on the magnetization through SOTs. So far, a number of studies on current-induced effective magnetic fields $H_{eff}$s have been reported in ferromagnetic semiconductor and metallic magnetic heterostructures. In ferromagnetic semiconductor GaMnAs, $H_{eff}$ has been attributed to strain-induced spin-orbit interaction in the bulk.[1] In metallic magnetic heterostructures, two main types of $H_{eff}$ have been reported: that induced by the Rashba–Edelstein effect at the interfaces,[2,6] and that induced by the spin Hall effect in a heavy metal layer adjacent to the ferromagnetic metal layer.[3] These metallic magnetic heterostructures are the basic structures forming practical magnetic tunnel junctions (MTJs), and three-terminal MTJs using $H_{eff}$s have been demonstrated.[3,7] Because three-terminal MTJs are expected to be the building blocks for next-generation electronics, thanks to their potential for operation at high speed with low power consumption, there has been intensive study of $H_{eff}$s in the metallic system. Half-metals enabling high-sensitivity detection of magnetization direction are promising because they can improve the performance of three-terminal MTJs. A half-metallic oxide $La_{0.67}Sr_{0.33}MnO_3$ (LSMO) is suitable for examining three-terminal MTJs because large tunnel magnetoresistance reflecting half-metallicity has been reported in a LSMO/SrTiO$_3$/LSMO MTJ,[8] though $H_{eff}$s have not yet been reported in it. In the present work, we investigate $H_{eff}$ in the simple half-metallic oxide heterostructure LSMO/SrTiO$_3$, using transport measurements to study $H_{eff}$s in oxide multilayers based on LSMO.

Here, LSMO films with nominal thickness of 13, 18, and 25 u.c. were grown on single-crystal (001) SrTiO$_3$ substrates by pulsed laser deposition, using a KrF excimer laser ($\lambda$ = 248 nm) to irradiate a sintered polycrystalline LSMO target at a repetition rate of 1 Hz. During growth, the substrate temperature and oxygen pressure were 750 °C and 25 Pa, respectively. After growth, the LSMO films were annealed for 20 min under conditions identical to the growth conditions, so as to recover their oxygen deficiency. Confirmed by X-ray diffraction and atomic force microscopy, all of the LSMO films were epitaxially grown on the substrate and showed atomically flat surfaces consisting of steps and terraces. The sheet resistance for each film, measured by the Van der Pauw method, increased with increasing temperature from 20 K, and its maximum sheet resistance was at 340–380 K, corresponding approximately to its Curie temperature.

The films were made into Hall bar devices with a 10-μm-wide channel along [100] and two pairs of Hall probes by conventional photolithography and wet etching, followed by lift-off of Au/Cr electrodes (Fig. 1(a)). Figure 1(b) shows a schematic of the measurement setup. The external magnetic field $H$ direction $\varphi$ and the magnetization $M$ direction $\varphi_M$ were measured from [110]. Since the equilibrium $M$ direction of LSMO is determined from the energy minimum condition under $H$, the magnetic anisotropy field $H_a$, and $H_{\text{eff}}$, we extracted $H_{\text{eff}}$ by analyzing how the $M$ direction depended on $H$ as a function of current $I$. The transverse resistance $R_{yx}$ of LSMO is dominated by the planar Hall effect and depends on the in-plane relative angle between the $I$ and the $M$ direction: $R_{yx} = R_{\text{PHE}} \sin(2\varphi_M + 90°) + R_{\text{off}}$, where $R_{\text{PHE}}$ is the amplitude of the planar Hall resistance and $R_{\text{off}}$ is the offset.[9] This method allowed us to monitor the $M$ direction by measuring $R_{yx}$. We also measured the longitudinal resistance $R_{xx}$ so we could calibrate the device temperature $T_d$, which increases by Joule heating under large $I$.[10] We evaluated $T_d$ by comparing $R_{xx}$ under $I$ with the temperature dependence of $R_{xx}$ measured at low current (50–100 μA).

We first examined the existence of $H_{\text{eff}}$ in the LSMO heterostructures. After aligning the $M$ direction by $\mu_0 H = 0.5$ T ($\mu_0$ is the permeability of vacuum) along [110], we measured $R_{yx}$ at $I = \pm 3$ mA while applying a rotating in-plane $H$ in the clockwise or counterclockwise direction. Figure 2(a) shows the $\varphi$ dependence of $R_{yx}$ for a 13-u.c. device under $\mu_0 H = 5$ mT at $T_d = 130$ K. Positive (negative) $I$ is defined as current flow from left (right) to right (left). Four hysteresis loops, which reflect $M$ switching across magnetic hard axes, appeared around the four-fold symmetry axes <100>, indicating that the LSMO had biaxial magnetic anisotropy and that its magnetic hard axes were nearly along <100>, as previously reported.[11] The hysteresis loops around [$\bar{1}$00] and [100] were wider than those around [010] and [0$\bar{1}$0]. Considering uniaxial magnetic anisotropy induced by step and exchange coupling with antiferromagnetic dead layer reported previously,[12,13] the present asymmetry could be caused by superposition of uniaxial magnetic anisotropy with the hard axes nearly along [100] ([$\bar{1}$00]) on the biaxial magnetic anisotropy. Figure 2(b)–(e) shows enlarged views around the four-fold symmetry axes <100>. The hysteresis loops around [010] and [0$\bar{1}$0] showed no dependence on current polarity, while the hysteresis loop around [$\bar{1}$00] ([100]) under positive (negative) $I$ shifted to larger $\varphi$ than that under negative (positive) $I$. From these results, we infer that the $H_{\text{eff}}$ along [010] ([0$\bar{1}$0]) was induced by positive (negative) $I$. Note that the contribution of the anomalous Nernst voltage ($\propto I \cos(\varphi_M + 45°)$) to $R_{yx}$ reported in GaMnAs[14] was evaluated to be less than 0.7% of $R_{yx}$ (within experimental error) from the difference in the amplitude of $R_{yx}$ under positive and negative $I$ in Fig. 2(a).

Next, we confirmed $H_{\text{eff}}$ by monitoring $M$ switching while applying $H$ parallel or antiparallel to $H_{\text{eff}}$. After aligning the $M$ direction by applying $\mu_0 H = 0.5$ T along [$\bar{1}$10] (nearly along one of magnetic easy axes), we measured $R_{yx}$ with various currents while sweeping the external magnetic field in [0$\bar{1}$0] (Fig. 3(a)). Figure 3(b) shows the loops of normalized $R_{yx}$ versus $\mu_0 H$ for a 13-u.c. device measured at $I = \pm 3$ mA and $T_d = 130$ K. A jump in the normalized $R_{yx}$, reflecting the switch in $M$ direction from [$\bar{1}$10] to [$\bar{1}\bar{1}$0], appeared at $\mu_0 H = 2.2$–2.5 mT. Intermediate values of $R_{yx}$ that appeared during switching may be related to nucleation of domains or to domain-wall motion. We define a switching magnetic field as a magnetic field giving a normalized $R_{yx} = 0$. The switching magnetic field under $I = 3$ mA is larger than that under $I = -3$ mA. The difference $H_{\text{diff}}$ between the switching magnetic fields under $\pm I$ is shown as a function of magnitude of $I$ in Fig. 3(c). Because half of $H_{diff}$ corresponds to $H_{eff}$ when coherent $M$ switching occurs, we reasonably suppose that $H_{diff}$ reflects the behavior of $H_{eff}$, even in the present case showing slight incoherent switching. The values of $H_{diff}$ are positive in the studied range of $I$, as expected from Fig. 2(d): $H_{eff}$ generated by positive (negative) current assists (hinders) the switching from $[\bar{1}10]$ to $[\bar{1}\bar{1}0]$. Moreover, the values of $H_{diff}$ increased with increasing magnitude of $I$, showing that $H_{eff}$ was induced by the application of $I$. We also measured the $T_d$ dependence of $H_{diff}$ under $|I|$ = 3 mA (Fig. 3(d)). The values of $H_{diff}$—that is, $H_{eff}$—increased with decreasing $T_d$, unlike an earlier result in GaMnAs.[15] The origin of the $T_d$ dependence of $H_{diff}$ in LSMO is currently unclear, and further systematic studies are necessary to clarify it.

Subsequently, we investigated $H_{eff}$ as a function of $I$ in three devices with different LSMO thicknesses. To prevent domain nucleation and domain-wall motion during $M$ switching, after applying $\mu_0 H$ = 0.5 T along $[\bar{1}10]$ at $T_d$ = 130 K, we measured the $\varphi$ dependence of $R_{yx}$ while rotating $H$ larger than $H_a$ around $[\bar{1}00]$ (one of the hard axes) counterclockwise, as previously reported.[1,15] To estimate $H_a$, we measured the $\varphi$ dependence of $R_{xx}$ (anisotropic magnetoresistance; AMR $\propto \cos(2\varphi_M + 90°)$) at low current (50–100 μA) while applying constant $H$. By calculating the dependence of $R_{xx}$ on $\varphi$ from the energy minimum condition under $H$ and $H_a$, including the biaxial anisotropy field $H_b$ and the uniaxial anisotropy field $H_u$, we performed the fit using $H_b$ and $H_u$ as fitting parameters, from which we obtained $H_a$. Figure 4(a) shows the $\varphi$ dependence of $R_{yx}$ for a 13-u.c. device while applying $\mu_0 H$ = 31 mT and $I$ = ±3 mA. We evaluated a switching angle where $M$ was directed to $[\bar{1}00]$ from the point giving the maximum of the curve's first derivative, considering the possible superposition of AMR on $R_{yx}$. The switching angle under positive $I$ was larger than that under negative $I$ by $H_{eff}$, consistent with Fig. 2(d). The displacement of the switching angle from 135° mainly comes from unintentional misalignment of the device. The magnitude of $H_{eff}$ is approximately $H_{eff} \approx H \sin(\Delta\varphi/2)$, where $\Delta\varphi$ is the difference in switching angles under $\pm I$.[1,15] To evaluate $H_{eff}$ under a certain $I$, we measured similar $R_{yx}$–$\varphi$ curves under constant $I$ while applying 3–6 different values of $H$ larger than $H_a$ for 3–6 times and averaged them for the applied $H$ and repetitions. Figure 4(b) shows the values of $H_{eff}$ at $T_d = 130$ K for devices with various LSMO thicknesses as a function of $I$. The error bars are determined from the standard deviations. For all the devices, $\mu_0 H_{eff}$ increased with increasing $I$, and thicker devices required a larger $I$ to generate a certain $H_{eff}$. Moreover, the values of $\mu_0 H_{eff}$ cannot be scaled with current density evaluated from the nominal LSMO thickness. Early experiments showed that ultra-thin LSMO (< 7 u.c.) was insulating.[16] Indeed, below 250 K, the resistivity of the present 6.5-u.c. LSMO was more than two orders of magnitude higher than that of the 13-u.c. LSMO. Assuming that a 6.5-u.c.-thick insulating layer is created in LSMO, the effective current density $J$ was evaluated from the effective LSMO thickness excluding the insulating layer. Figure 4(c) shows $H_{eff}$ as a function of $J$ at $T_d = 130$ K for devices with various LSMO thicknesses. The values of $H_{eff}$ were almost scaled with $J$, suggesting that $H_{eff}$ is induced in the LSMO bulk.

Finally, we compare the present results with mechanisms proposed for the current-induced effective magnetic field. The Oersted field generated from current flowing in the LSMO layer is not responsible for the observed $H_{eff}$ because $H_{diff}$ depended on $T_d$ (Fig. 3(d)) and the values of $H_{eff}$ were scaled with $J$ (Fig. 4(c)), whereas the Oersted field should be independent of temperature and dependent on $I$ rather than on $J$ in the present Hall bar geometry. We can also rule out current-induced domain-wall motion, whose direction is independent of the $M$ direction,[17-19] because the shift of the hysteresis loops in Fig. 2(d) and (e) depended on the $M$ direction. Considering the dependence of $H_{eff}$ on $J$ flowing in the conductive region of LSMO and the existence of the planar Hall effect, we believe that the observed $H_{eff}$ is related to spin-orbit interaction in LSMO and/or strain-induced spin-orbit interaction as reported in III–V semiconductors,[1,20] though clarifying the underlying mechanisms will require further theoretical support. Note that the efficiency of generation of $H_{eff}$ (ratio of $H_{eff}$ to $J$) in LSMO was evaluated as $6.4 \times 10^{-16}$ Tm$^2$A$^{-1}$ from the slope of the linear fit to the data in Fig. 4(c), which is two orders of magnitude smaller than that in ferromagnetic semiconductors.[1,15] This low efficiency may be related to weak spin-orbit interaction in LSMO compared with typical ferromagnets such as Ni.[21,22]

In summary, we investigated $H_{eff}$ in the half-metallic oxide LSMO by using transport measurements. Applying in-plane current to LSMO, an in-plane $H_{eff}$ orthogonal to the current direction appeared. $H_{eff}$ increased with increasing magnitude of $I$, indicating that $H_{eff}$ was induced by the applied in-plane $I$. The dependence of $H_{eff}$ on $I$ in devices with three LSMO thicknesses shows that the thicker device required a larger $I$ to generate a given $H_{eff}$. By assuming that a 6.5-u.c.-thick insulating layer is created in LSMO, we evaluated the effective current density. The values of $H_{eff}$ obtained at various LSMO thicknesses were almost scaled with the effective current density, suggesting that the observed $H_{eff}$ is induced in the LSMO bulk. Although improving the efficiency of $H_{eff}$ generation without degrading the half-metallicity is necessary for practical use, the $H_{eff}$ observed in this half-metal opens new perspectives for spintronics devices.

**Acknowledgements**

The authors thank J. Ieda for discussion. This work was supported by JSPS KAKENHI for Young Scientists A (15H05517) and Grant-in-Aid for Scientific Research on Innovative Areas (25106007) from the Japan Society for the Promotion of Science.

Figure captions

**Fig. 1.** (a) A micrograph of top view of a typical Hall bar device with a 10-μm wide channel along [100] and two pairs of Hall probes. (b) A schematic of measurement setup. Magnetic field direction $\varphi$ and magnetization direction $\varphi_M$ are measured from [110].

**Fig. 2.** (a) $\varphi$ dependence of normalized $R_{yx}$ for a 13 u.c. device measured with $\pm 3$ mA under application of rotating in-plane $\mu_0 H = 5$ mT in the clockwise and counterclockwise directions at $T_d = 130$ K. (b) and (c) ((d) and (e)) Enlarged views of (a) around $y$ ($x$) axis. The arrows show the rotation direction of $H$. Insets illustrate the corresponding switching of $M$ direction.

**Fig. 3.** (a) A schematic diagram of $M$ switching by $H$ along $[0\bar{1}0]$ under the application of $I$. (b) Normalized $R_{yx}$ versus $H$ curves for a 13-u.c. device under the application of $\pm 3$ mA at $T_d = 130$ K. The arrows show the sweep direction of $H$. (c) Dependence of difference $\mu_0 H_{\text{diff}}$ between switching magnetic fields under $\pm I$ on magnitude of $I$ at $T_d = 130$ K. (d) $T_d$ dependence of $\mu_0 H_{\text{diff}}$ under constant magnitude of $|I| = 3$mA.

**Fig. 4.** (a) $\varphi$ dependence of $R_{yx}$ for a 13-u.c. device in the application of $\mu_0 H = 31$ mT and $I = \pm 3$ mA at $T_d = 130$ K, where $H$ larger than $H_a$ is rotated across $[\bar{1}00]$ in the counterclockwise. (b) $I$ dependence of $\mu_0 H_{\text{eff}}$ evaluated from similar measurements to (a) at $T_d = 130$ K. (c) $\mu_0 H_{\text{eff}}$ shown in

(b) as a function of $J$ calculated by assuming that a 6.5-u.c.-thick insulating layer is created in LSMO.

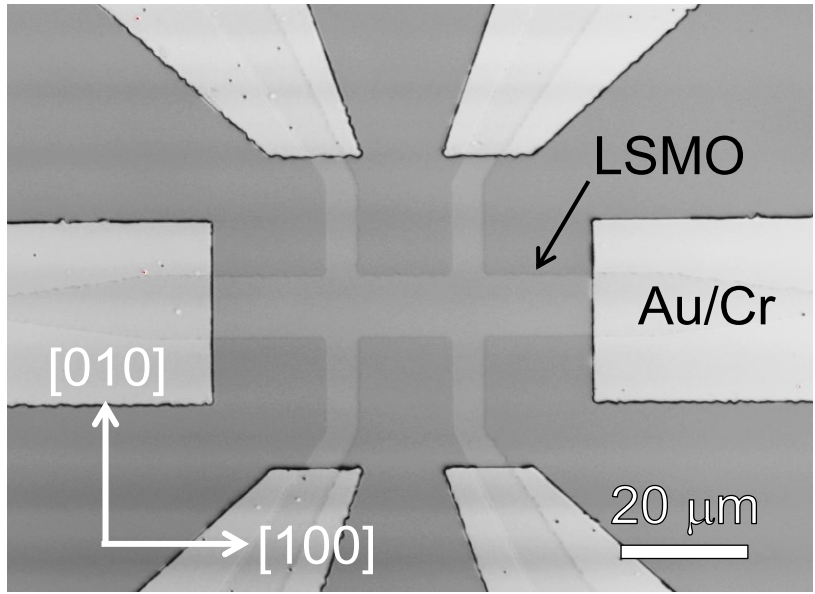 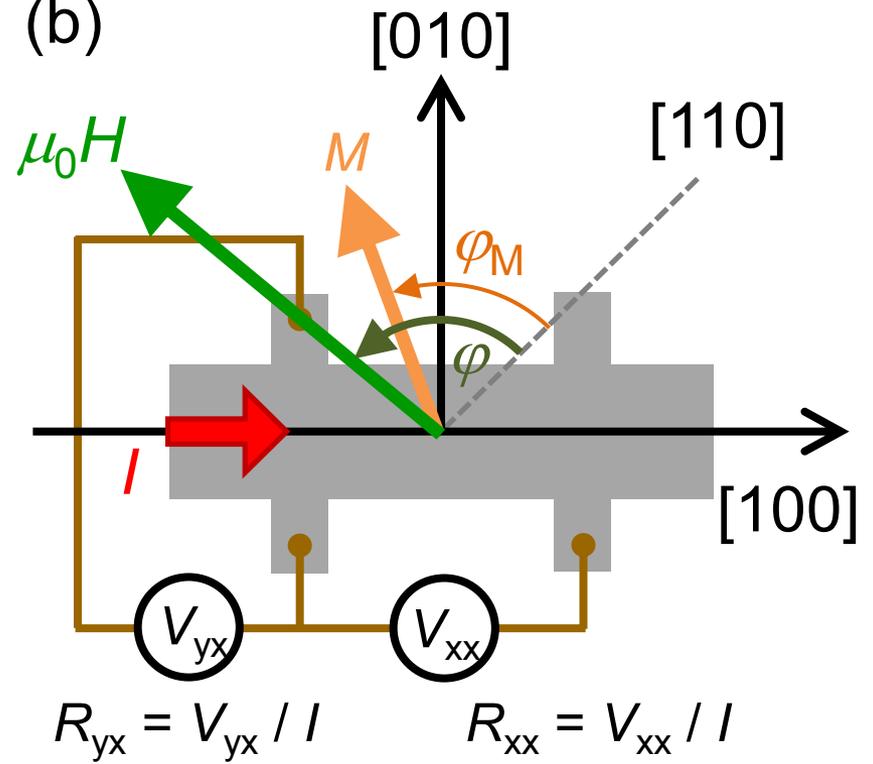

Fig. 1. Yamanouchi *et al*.

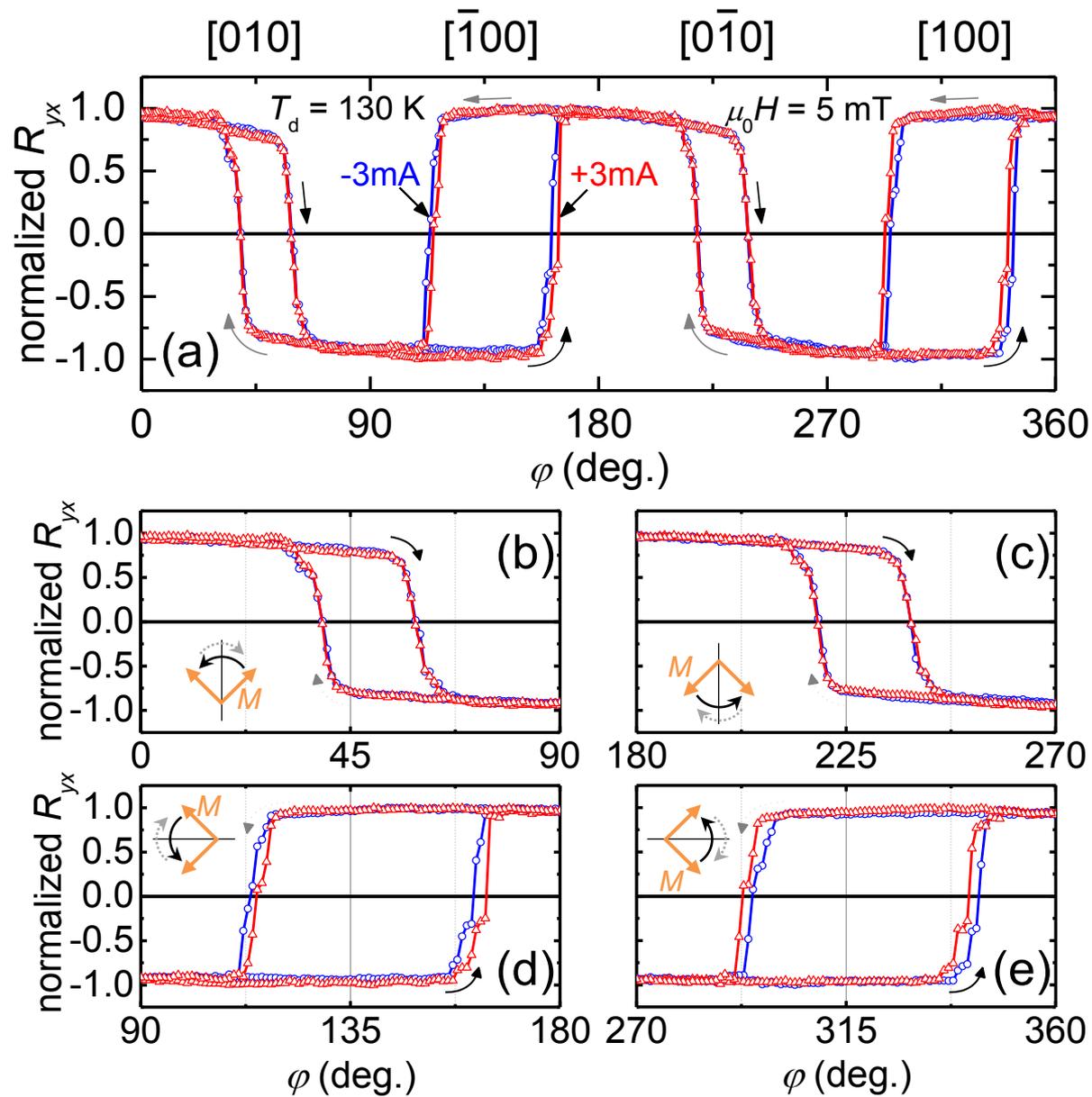

Fig. 2. Yamanouchi *et al*.

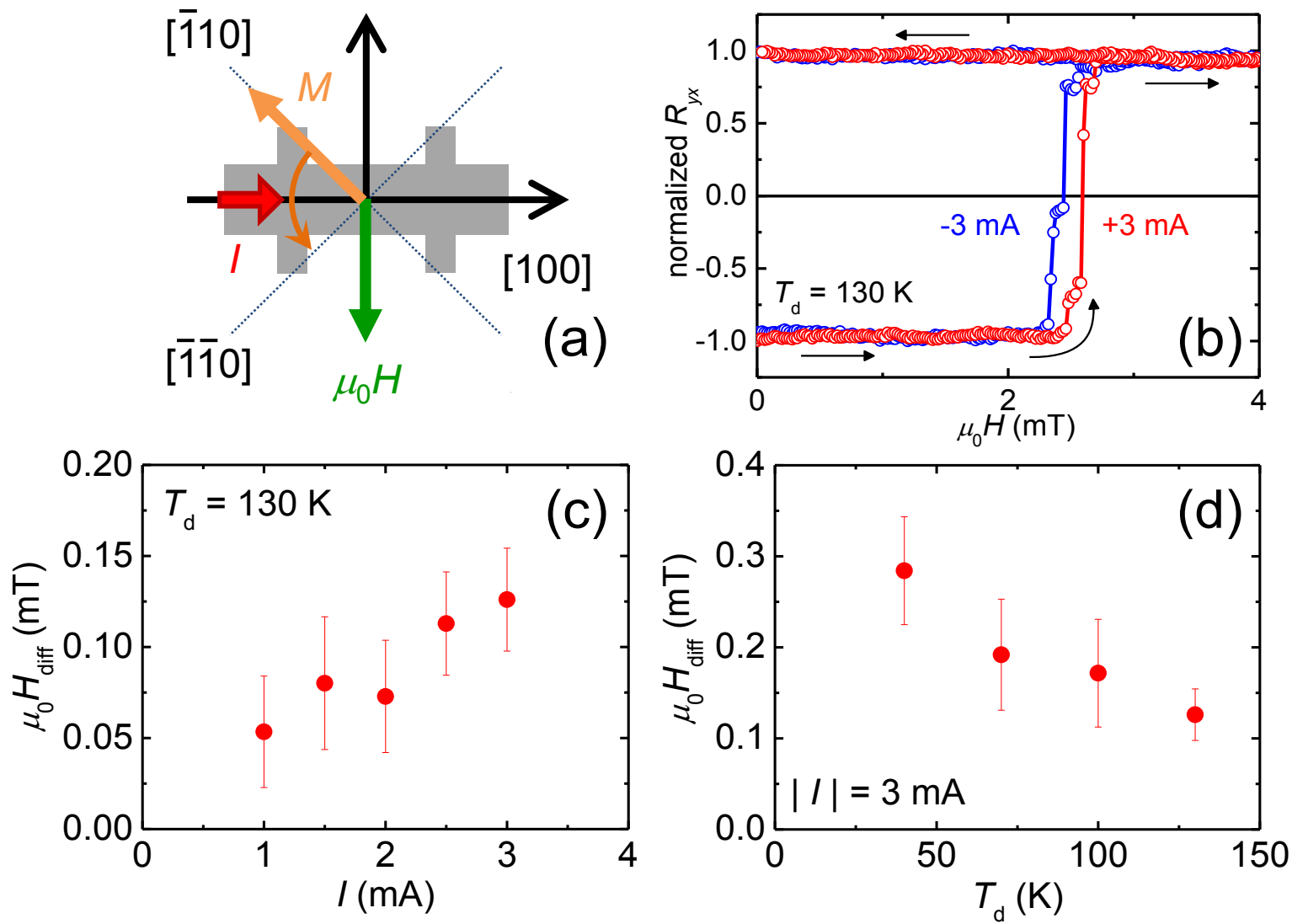

Fig. 3. Yamanouchi *et al*.

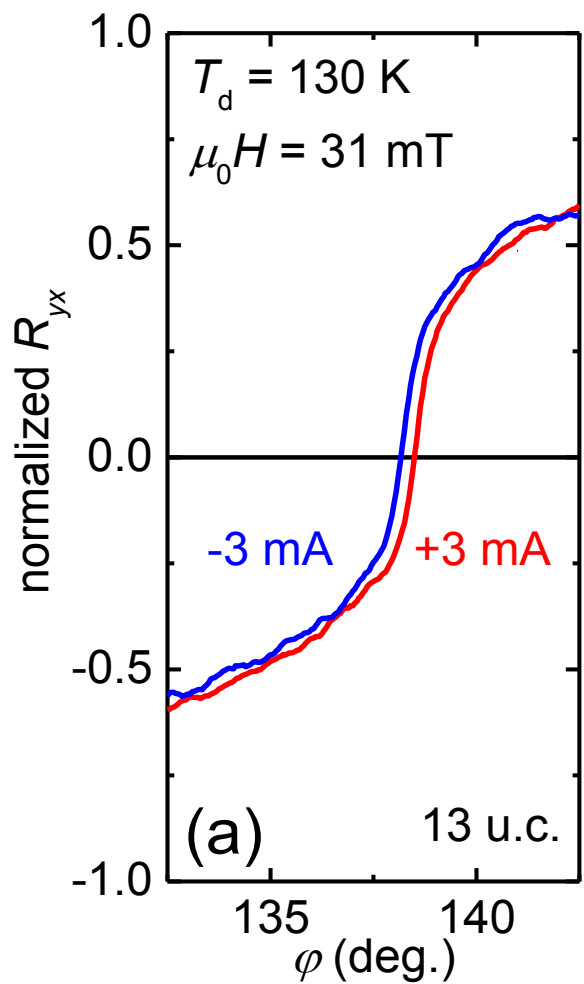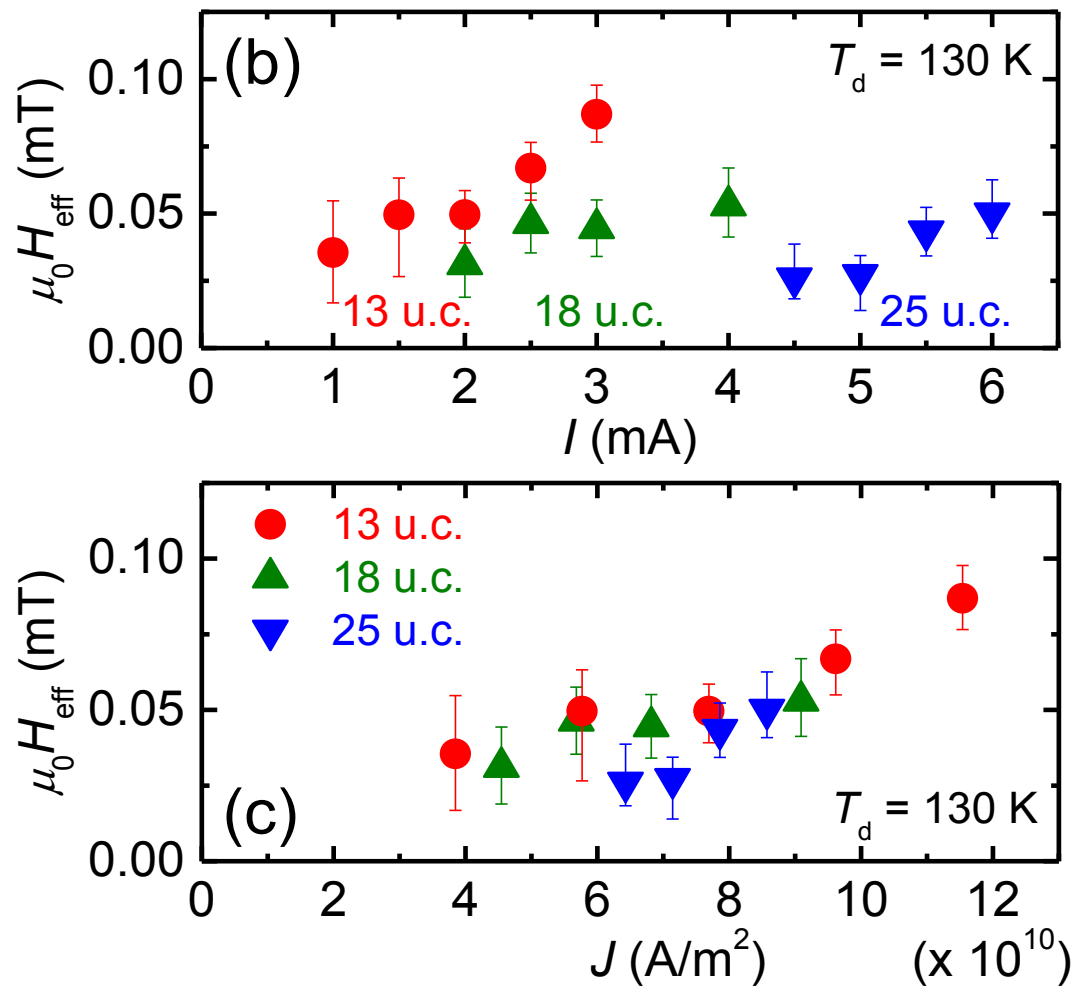

Fig. 4. Yamanouchi *et al*.